\documentclass[11pt]{article}
\usepackage{geometry}                
\usepackage{graphicx}
\usepackage{amssymb}
\usepackage{epstopdf}
\usepackage[OT2,OT1]{fontenc}
\newcommand\cyr{%
\renewcommand\rmdefault{wncyr}%
\renewcommand\sfdefault{wncyss}%
\renewcommand\encodingdefault{OT2}%
\normalfont
\selectfont}
\DeclareTextFontCommand{\textcyr}{\cyr}

\title{The probability for matter-antimatter segregation following the quark-hadron transition}
\author{Moishe Garfinkle\\Drexel University\\GarfinkM@Drexel.edu}

\begin{document}
\maketitle
 
\section{ABSTRACT}
Cosmologists such Sakharov, Alfv\`en, Klein, Weizs\"acker, Gamow and Harrison all disregarded the distribution of baryons and antibaryons immediately prior to freeze-out in trying to elucidate the circumstances that explained hadron distribution in the early universe.   They simply accepted a uniform distribution: each baryon paired with an antibaryon. The acceptance of this assumption lead to many problems they were unable to overcome.  This essay discards this assumption of homogeneity or uniformity.  Although this essay does deal with early-universe matters, it is not meant to indicate any involvement in energy distribution functions nor in any symmetry-asymmetry controversies.  Cluster formation is strictly geometric.   This essay has value as far as problems early cosmologists faced but also should complete the historic record. 

\section{Introduction}
The Standard Cosmological Model (SCM) has so well integrated celestial and relativistic mechanics into a coherent description of the universe that it is a little less than astonishing just how well the SCM is in agreement with empirical observations.   Though some minor problems still persist these involve incomplete concepts rather than any gross errors.  These include galaxy and supercluster formation despite the near-uniformity of the cosmic background radiation\cite{deOlive} and the horizon problem wherein the observed thermal equilibrium of distant portions of the universe that could not have been in contact in the time allotted since the formation of the universe\cite{Levi}.   

The major discordant feature to this largely idyllic picture involves the ongoing debate between proponents of a symmetric universe comprising absolutely equal amounts of matter and antimatter\cite{Glashow} and proponents of an asymmetric universe comprising essentially of matter alone\cite{Jacobs}.   The roots of this controversy date back to 1928. From the quantum mechanical concept of time-reversal symmetry Dirac predicted that there exists for each matter particle an antimatter equivalent\cite{Dirac}.    Anderson observed particles with same mass as the electron but with opposite charge, thus confirming Dirac's prediction\cite{Anderson}.  Within the next ten years all of the antiparticles of the principal hadrons had been discovered.   After the discovery of antimatter equivalents to matter particles the symmetry of the universe was taken for granted: for every matter particle there existed an antimatter particle.

The particles of concern in this discourse are the principle nuclear constituents: the baryons and antibaryons.   These are comprised of various combinations of primordial particles called quarks.   Quarks are elementary particles and are the fundamental constituent of matter.    The usual baryons and antibaryons are composed of quark triplets and comprise the visible universe of ordinary matter\cite{Nave}.

Unfortunately, according to the SCM spontaneous annihilation of these baryon and antibaryon pairs in a Symmetric Universe would have occurred at freeze-out at which temperature the rate of baryon and antibaryon pair creation fell to zero while their annihilation accelerated\cite{Peebles}.   Many attempts by various theoreticians have been made to no avail to save the Symmetric Universe by averting this freeze-out catastrophe, as will be discussed presently.

Alternatively, if it were assumed that antibaryons were absent in the early universe then this annihilation problem could be avoided: the universe would largely comprise baryons.  Unfortunately, either no means have been proposed, or too many discordant proposals have been advanced\cite{Wands}, to effect the elimination of antibaryons in a manner consistent with the Standard Particle Model (SPM) which relates three fundamental interactions with elementary\cite{Glashow2}.   The SPM ostensibly requires that baryon number is maintained, though this last requirement is being challenged\cite{Allday}.

The objective of this study is simply to present an alternative model of baryon-antibaryon segregation at freeze-out consistent with both the SCM and the SPM without taking sides in this controversy between proponents of a symmetric versus the proponents of an asymmetric universe. 

Essentially, proponents of a symmetric universe in which baryons exactly equal antibaryons ($B=0$) is in agreement with the SPM in which all known reactions (with several minor exceptions) maintain matter-antimatter symmetry.  According to the SCM however, following the quark-hadron transition at $T \ll 1$ TeV baryons and antibaryons were in thermal equilibrium with radiation: annihilation rates equaled creation rates\cite{Kolb}.  This equilibrium persisted until the universe cooled to $T \approx 20$ MeV: freeze-out, at which temperature equilibrium was broken: baryons and antibaryons annihilation was not compensated by baryons and antibaryons creation\cite{Applegate}.

Accordingly, the annihilation process would be so complete that only a miniscule number of baryons and antibaryons would have survived freeze-out:  roughly $n_B / n_\gamma = n_{\overline{B}} / n_\gamma  \approx 10^{Ð18}$ rather than the expected $n_B / n_\gamma = n_{\overline{B}} / n_\gamma \approx 10^{Ð10}$.    The present universe could not have survived freeze-out, nor could the SCM.

To ensure the survival of the SCM a means had to be found to circumvent annihilation at freeze-out.    All of the investigators involved accepted essentially without thought that the distribution of particles before freeze-out was uniform: every baryon spatially paired with every antibaryon.   In this regard the most promising possibility was macroscopic separation of baryons from antibaryons before annihilation.
 
Alfv\'en \cite{Alfven}, and Klein\cite{Klein} among others, have proposed symmetric cosmological mechanisms that unfortunately collapse at freeze-out because the matter-antimatter segregation processes projected proved ineffectual. Weizs\"acker\cite{Weizsacker} proposed turbulence to promote condensation that Gamow\cite{Gamow} \em et al. \rm applied to perturbations that might lead to matter-antimatter segregation and failed even worse.  Using Hubble expansion, Harrison tried gravitational perturbations and failed again\cite{Harrison}.

The failure of any of the separation processes to achieve any positive results led Sakharov as a last resort to propose an asymmetric universe ($B \rightarrow 1$)\cite{Sakharov}. He hypothesized axiomatically that:

\begin{quote}\begin{em}``The theory of the expanding Universe, which presupposes a superdense initial state of matter, apparently excludes the possibility of macroscopic separation of matter from antimatter.  It must therefore be assumed that there are no antimatter bodies in nature, i.e. the Universe is asymmetrical with respect to the number of particles and antiparticlesÓ. \end{em} \end{quote}

This was a bold declaration made to save the SCM.   Sakharov proposed this hypothesis without any reliance whatsoever on any empirical evidence\cite{Sakharov}. It is now well over a half-century since this radical solution was proposed, and as of yet no definitive means have appeared to achieve this end: a baryon-antibaryon asymmetry persisting after the quark-hadron transition.  

For example, by assuming that the number of baryons exceeded the number of antibaryons by some amount following the quark-hadron transition then on freeze-out all of the baryons-antibaryon pairs would annihilate, leaving the unpaired baryons.  The number of baryons necessary to survive freeze-out was calculated by Kuzmin \em et al. \rm \cite{Kuzmin} and found to be only $n_B \approx (10^9+1)/10^9$: essentially one in a billion.

Regardless of how small a proportion of baryons need survive, how this asymmetry arose still needed to be explained.   Grand unification theories, both equilibrium and non-equilibrium have been resorted to with disappointing results, as has baryogenesis based on grand unification considerations\cite{Narlikar}   Resort has been made to CP violations in certain mesons.  These ostensibly neutral mesons can exist in either matter or antimatter states as observed in $K^0$ and $B^0$ mesons, with matter favored over antimatter. 

If such an unequal distribution favored matter particles over antimatter particles in the early universe then the possibility exists that matter particles sufficiently outnumbered antimatter particles and that this discrepancy survived the quark-hadron transition.   At freeze-out the excess matter particles would have survived annihilation to form our present asymmetric universe matter-dominated universe. This scenario is a logical construct, but CP violations in mesons do not as yet have a modicum of cosmological significance.

Attempts have been made to compute the baryonic asymmetry from first principals using different methodologies by Goa et al.\cite{Goa}, Dolgov\cite{Dolgov}, and Dudarewicz et al.\cite{Dude} among others, again without substantial results.    In general observational evidence for a universal baryon asymmetry is weak Cohen et al. \cite{Cohen}  Sakharov's\cite{Sakharov} hypothesis is still gaining adherents but still without definitive empirical substantiation, more than a half-century after he proposed it. 

According to Ting, the prospective observations to be made with the Alpha Magnetic Spectrometer (AMS) might answer the asymmetry question \cite{Ting}.    Battiston confirms they will be looking for interstellar antimatter particles\cite{Battiston}.   However interpretation of the results will be a major concern because many interoperations of the data are possible.   Consequently, there is some doubt as to how definitive the result might be and how to properly decipher them\cite{Bottino}.

Evidently the problem is not the paucity of possible solutions, but the overwhelming number proposed.    Consider now an alternative approach based on a particle distribution function.  Energy distribution between particles would play no role in this spatial distribution of the particles themselves because the distribution is purely geometric.   This essay is  presented solely to indicate that there exists an alternative mechanism for matter-antimatter segregation at freeze-out than those already quoted.

\section{Distribution}

Consider again the nature of the primordial universe just after the quark-hadron transition at $T \ll 1$TeV.   At this point the universe was one in which baryons and mesons and their anti-equivalents were in thermal equilibrium with radiation: particle creation processes were in balance with annihilation processes\cite{Peebles}.  As the temperature decreased,  annihilation processes began to dominate, with creation processes continually diminishing\cite{Rich}. These creation processes essentially ceased at $T \approx 20$ MeV: denoted ``freeze-out"\cite{Peebles}.  The creation of matter was no long able to keep up with its annihilation.

Alfv\'en \cite{Alfven}, Klein\cite{Klein}, WeizsŠcker\cite{Weizsacker}, Gamow\cite{Gamow}, Harrison\cite{Harrison}, and Sakharov\cite{Sakharov} all disregarded the distribution of baryons and antibaryons immediately prior to freeze-out, simply accepting a uniform distribution: each baryon paired with an antibaryon. Disposing of this assumption, consider now a probabilistic approach based on a binomial distribution of baryons and antibaryons\cite{Mendenhall}.

The spatial distribution of particles can be modeled as a linear sequence as demonstrated by Cantor \cite{Cantor}.   As an example consider a large volume containing discrete particles equally divided between baryons and antibaryons.  We will now start removing a single particle at a time from the volume according to the Binomial formulation to determine whether it is a baryon or antibaryon.

\begin{enumerate}
\item Each trial has one possible outcome: either success or failure.

\item The trials are independent. Thus, the outcome of one trial has no influence over the outcome of another trial. Arguably the probability has changed with the removal of a single particle, but the number of available particles is so large that the change is insignificant to this pursuant discussion.

\item The total number of trials is limited only by the number of particles.

\end{enumerate}

A baryon will be considered a successful draw for each trial.   If the particles are perfectly arranged (every baryon is spatially paired with an antibaryon) then in some finite sub-set of particles, the number of baryons will equal the number of antibaryons.  As the size of this subset decreases, the degree of order increases.   In this case we should quickly reach essentially a 1/2 probability of picking baryons.   In fact the chance of such a perfect distribution is very slim.  The binomial distribution takes the form:

\begin{equation}
p(k)   =   C(n;k) p^k q^{n-k}
\label{eq1}
\end{equation}

where $p$ is the probability of a successful draw (a baryon), $q$ is the probability of an unsuccessful draw (an antibaryon), $n$ is the total number of particles, $k$ the number drawn, and $C(n;k)$ is the binomial coefficient

\begin{equation}
C(n;k)  = \frac{ n!}{k! (n-k)!}
\label{eq2}
\end{equation}

	For example consider the case for which $p=1/2$ and therefore $q=1/2$, the ideal arrangement for $1,000$ baryons and antibaryons arrayed in a $10 \times 10 \times 10$ cubic space.    The binomial distribution for $n = 1,000$ baryons and antibaryons after $k = 500$ draws is only $p(k) = 1/40$. Accordingly, the probability of a uniform distribution of just $1,000$ particles is less than one-tenth the expected $p = 1/2$.  The distribution for $n = 10,000$ baryons after $k = 5,000$ draws is down to $p(k) = 1/125$.  For $n = 100,000$ particles after $k = 50,000$ draws: $p(k) = 1/1,250$.  Again ideally it should be roughly $p(k) = 1/2$ for an ideal or uniform distribution.    

As the number of particles $n$ increases to cosmological proportions the probability for uniform distribution rapidly diminishes, and is infinitesimally small for the actual number of baryons present at freeze-out.  The expected ideal arrangement is probabilistically unrealistic.  Essentially the probability for all of the baryons and antibaryons to be exactly paired prior to freeze-out is nil.   Rather the most probable case is misplaced baryons and antibaryons.  That is: baryon clusters and antibaryon clusters of varying sizes are prominent in the particle distribution, in agreement with this probabilistic analysis. 

\begin{figure}
\begin{center}
\includegraphics[scale=0.75]{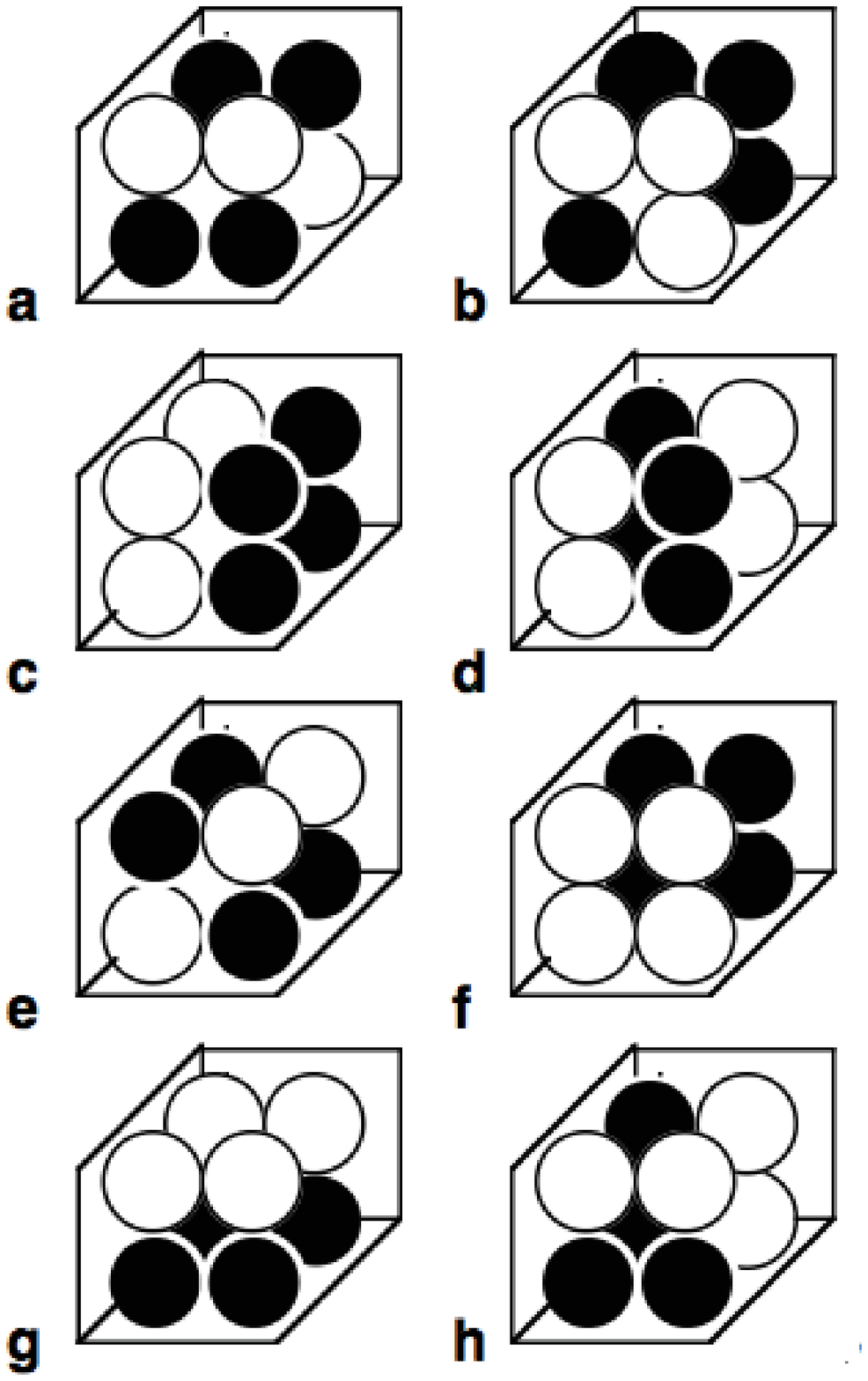}
\end{center}

\caption{Eight balls in several cubic arrangements}
\label{fig1}
\end{figure}

Consider now a box containing eight balls, half baryons (white) and half antibaryons (black) in a simple $2 \times 2 \times 2$ cubic array.    Eight of the roughly $64$ arrangements possible are illustrated in Figure \ref{fig1}.    Only six of these arrangements involve perfect segregation: Figures 1c, 1f, 1g and their mirror images.

On vigorously shaking the box the probability of achieving segregation would be only $0.094$, less that 10\%.

Consider again a box, but now containing 1,000 particles, half baryons and half antibaryons, in a simple $10 \times 10 \times 10$ cubic array.    Of the roughly $10^{300}$ particle arrangements possible, again only six would show perfect segregation: each a mirror image of the other.   Rather cluster sizes can range from 2-particle clusters to a 500-particle cluster: perfect segregation. 

\begin{figure}
\begin{center}
\includegraphics[width=6in]{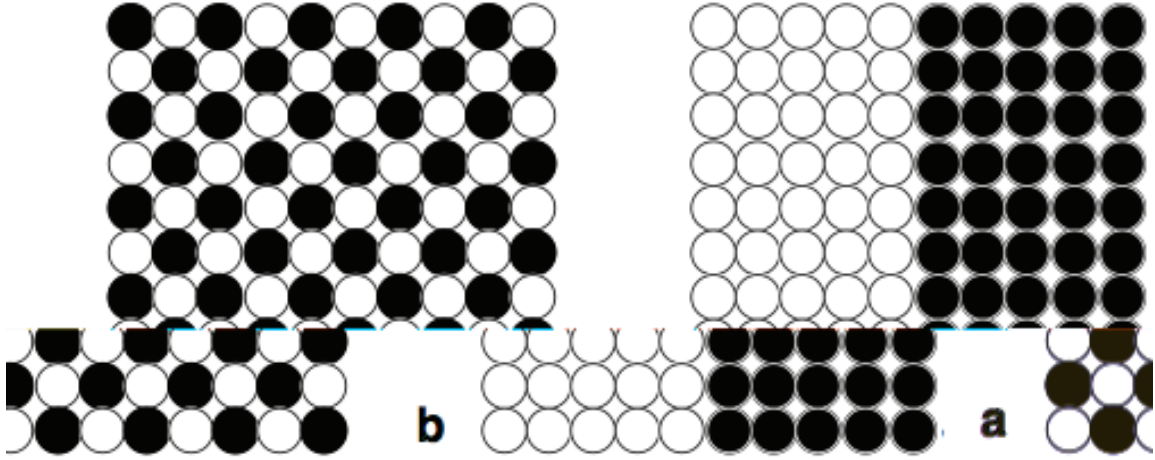}
\caption{A 10 $\times$ 10 array of balls in two cubic arrangements}
\end{center}
\label{fig2}
\end{figure}

 For example, Figure \ref{fig2}a illustrates a perfect symmetric arrangement, the probability of which is essentially nil.   Figure \ref{fig2}b illustrates perfect segregation, of which there are four arrangements, including their mirror images.   Out of the number of possible arrangements possible, the probability of perfect segregation is again essentially nil.

This discussion is not to suggest that this example \begin{em}per se\end{em} has cosmological significance but to illustrate that both the geometric consequences of clustering and its general agreement with the probabilistic analysis: as the number of particles increases the number of possible cluster arrangements increases more rapidly.   Of course this example is an idealized static simplification but the possibilities are not.  Probabilistically, the actual particle distribution prior to freeze-out is not uniformity but involving discrete baryon and antibaryon clusters.   These clusters will be disintegrating and reforming between the quark-hadron transition and freeze-out, but probability for clustering will be essentially unaltered.   For the cosmological number of particles at freeze-out the appearance of clustering is the rule.

\begin{figure}
\begin{center}
\includegraphics[width=6in]{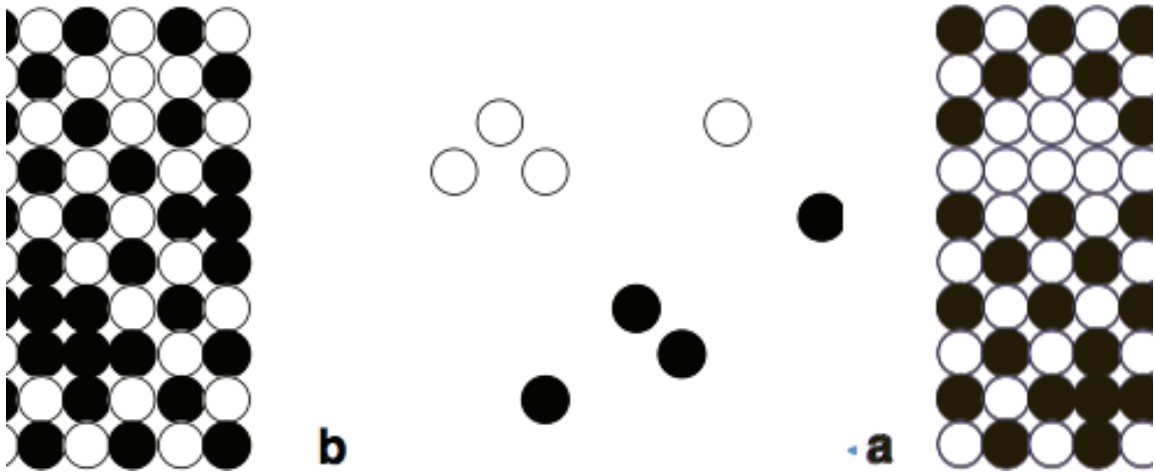}
\end{center}

\caption{A $10 \times 10$ array of balls before and after annihilation}
\label{fig3}
\end{figure}

Figure \ref{fig3}a illustrates several misplaced particles against a background of particle uniformity.  If the background particles are removed (annihilated) as shown in Figure \ref{fig3}b, then only the misplaced particles survive.    

Only those baryons paired with antibaryons will immediately annihilate at freeze-out.   Hence freeze-out from strictly a thermodynamic (energetic) viewpoint will result in baryon mass extinction, but from the more realistic kinetic (mechanistic) viewpoint this is hardly the case.   Freeze-out is now a time-dependent process: as the thermodynamics for annihilation becomes more favorable with decreasing temperature the kinetics becomes less favorable, finally ceasing altogether. So what can one presume?

The particle population will comprise both adjacent $n_B$ pairs and unpaired or clustered baryons and antibaryons.    Adjacent pairs will first annihilate and then the baryons or antibaryons clusters.   The closest of these baryons and antibaryons clusters will then annihilate leaving the remainder the larger baryons or antibaryons clusters.  Before it is thought that this process simply continues recall that freeze-out is now a time-dependent occurrence only initiating at $T = 20$ MeV and subsequently controlled by kinetic processes.  Because Hubble expansion will have increased inter-cluster distances some ten-fold between freeze-out and nucleosynthesis, annihilation is marginally less certain.   Nevertheless, annihilation is still the overwhelmingly predominant process involved, but recall: only one particle out of every billion particles annihilated need survive to meet the Kuzmin criteria.  What is proposed is no more than an essentially insignificant shift in the annihilation process at freeze-out, but possibly fundamental to the very survival of a symmetric universe.

Between freeze-out and nucleosynthesis non-Hubble drift will bring particles in contact, with both matter and antimatter clusters growing.   However contact of baryon and antibaryon clusters will not necessarily result in annihilation.  Only annihilation of surface particles in baryon and antibaryon clusters will occur, causing cluster recoil.  Hence like-clusters will agglomerate on contact while unlike-clusters will be driven apart. 

\section{Conclusion}
The great number of possibilities advanced to prevent universal annihilation at freeze-out are all based on the acceptance of a uniform distribution of baryons and antibaryon, an eventuality with an infinitesimally small possibility. Cluster formation, rather than being an anomaly, is the expected distribution, and not dependent on particle energy distribution.  This conclusion is not intended to enter into the ongoing debate concerning the symmetric versus asymmetric universe but simple to include a hitherto unrealized possible arrangement of particles at the quark-hadron transitions to complete the historic picture.

\end{document}